\documentclass[aps,prc,onecolumn,superscriptaddress,showpacs,showkeys]{revtex4}
\usepackage{amsfonts,amssymb,amscd,amsmath}
\usepackage{graphicx}
\usepackage{subfigure}
\usepackage{appendix}
\graphicspath{{figures/}}

\newcommand{\rr}{\mbox{\boldmath $r$}}
\newcommand{\rp}{\mbox{\boldmath $r^{\prime}$}}
\newcommand{\ba}{\mbox{\boldmath $a$}}
\newcommand{\bE}{\mbox{\boldmath $E$}}
\newcommand{\be}{\mbox{\boldmath $e$}}
\newcommand{\bj}{\mbox{\boldmath $j$}}
\newcommand{\bsigma}{\mbox{\boldmath $\sigma$}}
\newcommand{\nv}{\mbox{\boldmath $n$}}
\newcommand{\pv}{\mbox{\boldmath $p$}}
\newcommand{\lv}{\mbox{\boldmath $l$}}
\newcommand{\kv}{\mbox{\boldmath $k$}}
\newcommand{\bv}{\mbox{\boldmath $b$}}
\newcommand{\sv}{\mbox{\boldmath $s$}}
\newcommand{\Bv}{\mbox{\boldmath $B$}}
\newcommand{\Deltav}{\mbox{\boldmath $\Delta$}}

\newcommand{\ket}[1]{{#1} \rangle}
\newcommand{\bra}[1]{\langle {#1} }

\newcommand{\Bigket}[1]{{#1} \Big\rangle}
\newcommand{\Bigbra}[1]{\Big\langle {#1} }

\newcommand{\dd}{\, \mathrm{d}}
\newcommand{\rb}{\mbox{\boldmath $b$}}
\begin{document}
\title{Production of electroweak gauge bosons at forward rapidities  \\ in the  color -  dipole $S$ - matrix  framework}

\author{Yan B. {\sc Bandeira}}
\email{yanbuenobandeira@gmail.com }
\affiliation{Institute of Physics and Mathematics, Federal University of Pelotas, \\
  Postal Code 354,  96010-900, Pelotas, RS, Brazil}
\affiliation{Institute of Nuclear Physics PAN, PL-31-342 Cracow, Poland}

\author{Victor P. {\sc Gon\c{c}alves}}
\email{barros@ufpel.edu.br}
\affiliation{Institute of Physics and Mathematics, Federal University of Pelotas, \\
  Postal Code 354,  96010-900, Pelotas, RS, Brazil}
\affiliation{Institute of Modern Physics, Chinese Academy of Sciences,
  Lanzhou 730000, China}

\author{Wolfgang {\sc Sch\"afer}}
\email{Wolfgang.Schafer@ifj.edu.pl}
\affiliation{Institute of Nuclear Physics PAN, PL-31-342 Cracow, Poland}

\begin{abstract}
The cross-section for the production of an electroweak gauge boson ($G = W^{\pm}, Z^0,  \gamma$)   at forward rapidities in $pp$ collisions is derived within the  color -  dipole $S$ - matrix framework. We present the full expressions for the differential cross-section of the $q p \rightarrow G X$ process in the impact parameter  and  transverse momentum spaces, considering the longitudinal and transverse polarizations of the gauge boson. The particular cases associated with the  Drell - Yan process and
real photon production are discussed. We demonstrate that the final formulae are expressed in terms of the   dipole - proton cross-section or the unintegrated gluon distribution,  and can be used to estimate the impact of the saturation effects in the gauge boson production at the LHC and future colliders.
\end{abstract}
\maketitle

\section{Introduction}
The particle production at forward rapidities in hadronic collisions is  one  of the most promising processes to probe the QCD dynamics at small - $x$ as well as to observe the  breakdown of the collinear and $k_T$ factorization theorems, predicted to occur to high partonic densities \cite{hdqcd}. 
In this process, one has the interaction between projectile partons with large light cone momentum fractions ($x_1 \rightarrow 1$) and target partons carrying a very small momentum fraction
($x_2 \ll 1$). Consequently,  the partons from the projectile scatter off a dense gluonic system in the target and are expected to undergo multiple scatterings, which cannot be easily encoded in
the traditional (collinear and $k_T$) factorization schemes, in particular for a nuclear target.  
In recent years, several groups have proposed generalized factorization schemes, which involve new objects in addition to the usual (un)integrated parton distribution functions \cite{Nikolaev:2003zf,Nikolaev:2004cu,Nikolaev:2005dd,Nikolaev:2005zj,Nikolaev:2005ay,Nikolaev:2005qs,Fujii:2005vj,Dominguez:2011wm,Dominguez:2012ad,Kotko:2015ura}. In particular, in the hybrid factorization formalism,  the hadronic cross-section for the dihadron production at forward rapidities is schematically expressed as follows (For a recent review, see e.g. Ref. \cite{vanHameren:2023oiq}):
\begin{eqnarray}
d\sigma(h_A h_B \rightarrow H_1 H_2 X) \propto  f_{a/A}(x_1) \otimes d\sigma(a B \rightarrow b c) \otimes D_{H_1/b} \otimes D_{H_2/c} \,\,,
\end{eqnarray}
i.e., as a convolution of the standard
parton distributions for the dilute projectile, the parton - target cross-section (which includes the high-density
effects) and the parton fragmentation functions. As demonstrated, e.g., in Refs.  \cite{Nikolaev:2003zf,Dominguez:2012ad}, the derivation of $d\sigma(a B \rightarrow b c)$ for the case where the final state particles are partons is not a trivial task, and implies e.g. the presence of quadrupole correlators of fundamental Wilson lines or the contribution of a four partons $S$ matrix, depending on the approach considered estimating this quantity. The calculation of these objects in the general case  is still a challenge.

\begin{figure}[t]
{\includegraphics[width=0.6\textwidth]{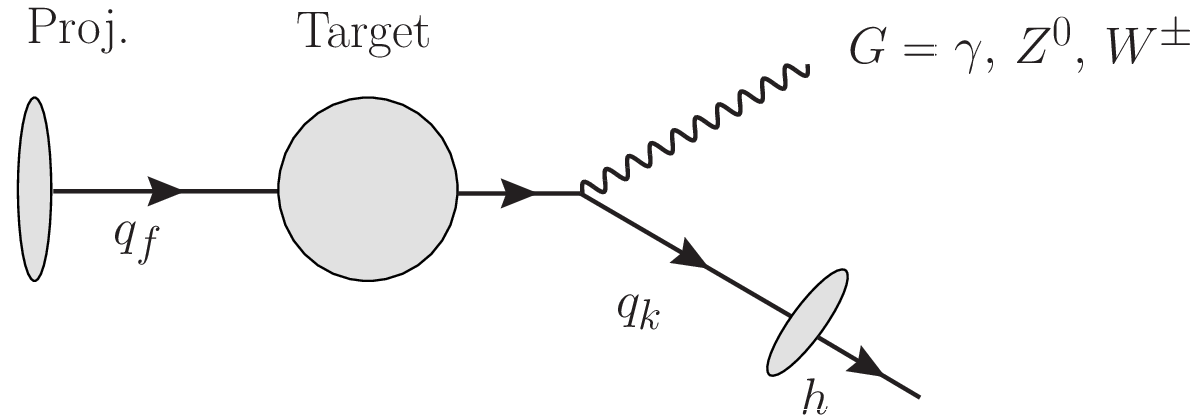}}  
\caption{Typical diagram contributing for the gauge boson + hadron  production in hadronic collisions, where the  gauge boson is irradiated by a quark  of flavor $f$  after  the interaction with the target color field (denoted by a shaded circle). For the $W^{\pm}$ radiation one has $q_k \neq q_f$.}
\label{fig:diagram}
\end{figure}

In this paper, we will consider one simplest case, where one of the particles in the final state is an electroweak gauge boson ($G = W^{\pm}, Z^0,  \gamma$), as represented in Fig. \ref{fig:diagram},  and will restrict our study to the isolated gauge boson production in $pp$ collisions, postponing the treatment of the associated gauge boson + hadron production and the generalization for a nuclear target for forthcoming publications. Our analysis is motivated by the studies performed in Refs. \cite{Kopeliovich:2007yva,Kopeliovich:2009yw,SampaiodosSantos:2020lte,Gelis:2002ki,Jalilian-Marian:2012wwi,Ducloue:2017kkq,Goncalves:2020tvh,Lima:2023dqw,Kopeliovich:2000fb,Raufeisen:2002zp,Kopeliovich:2001hf,Betemps:2004xr,Betemps:2003je,Golec-Biernat:2010dup,Ducati:2013cga,Schafer:2016qmk,Ducloue:2017zfd,Gelis:2002fw,Baier:2004tj,Stasto:2012ru,Kang:2012vm,Basso:2016ulb,Basso:2015pba}, which have estimated the cross-sections for the Drell - Yan, real photon and $Z^0$ production in $pp$ collisions at forward rapidities, using distinct approaches and assuming different approximations in the calculation of the $q p \rightarrow G X$ cross-section. In this paper, we will derive in detail the general expression for $\sigma(q p \rightarrow G X)$ and will demonstrate that it reduces to those used in Refs. \cite{Kopeliovich:2007yva,Kopeliovich:2009yw,SampaiodosSantos:2020lte,Gelis:2002ki,Jalilian-Marian:2012wwi,Ducloue:2017kkq,Goncalves:2020tvh,Lima:2023dqw,Kopeliovich:2000fb,Raufeisen:2002zp,Kopeliovich:2001hf,Betemps:2004xr,Betemps:2003je,Golec-Biernat:2010dup,Ducati:2013cga,Schafer:2016qmk,Ducloue:2017zfd,Gelis:2002fw,Baier:2004tj,Stasto:2012ru,Kang:2012vm,Basso:2016ulb,Basso:2015pba}  in the appropriate limits and representations.  Moreover, we will derive, for the first time, the cross-section for the $W^{\pm}$ production, which can be used to estimate the production of this final state in the kinematical range probed by the LHCb detector. 

Our calculations will be performed using the color - dipole $S$ - matrix framework for hadronic collisions, proposed in Refs. \cite{Nikolaev:1994de,Nikolaev:1995ty} and generalized in a series of publications \cite{Nikolaev:2003zf,Nikolaev:2004cu,Nikolaev:2005dd,Nikolaev:2005zj,Nikolaev:2005ay,Nikolaev:2005qs} (See also Refs. \cite{k95,bhq97,kst99,Kopeliovich:2002yv} for a related approach). Such an approach will be briefly reviewed in the next section, and the master equation derived in Ref. \cite{Nikolaev:2003zf} for a generic final state will be applied for the electroweak gauge boson production. As we will show, one of the main inputs for the calculation of  $\sigma(q p \rightarrow G X)$ in this approach is the 
light cone wave function (LCWF) of the gauge boson radiation off a quark, which will be derived in detail in section \ref{sec:wf} keeping the quark masses. In particular, these results can be applied to the $q_a \rightarrow q_b W^{\pm}$ transition, with the associated expression being  presented here for the first time. In section
\ref{sec:representations}, we will use the results for the LCWF's to derive the differential cross-sections for the isolated gauge boson production in the impact parameter and  transverse momentum spaces, which are expressed in terms of the dipole - proton cross - section and unintegrated gluon distribution of the proton, respectively. The vector and axial contributions will be estimated, and the expressions for the longitudinal and tranverse polarizations will be explicitly presented. In section \ref{sec:representations}, we will discuss some particular cases of our general expression and demonstrate that it reduces, in some specific limits, to the cross-sections used in the literature to calculate the Drell - Yan process and the real photon production. In addition, we will estimate the spectrum in the color transparency limit of the dipole - proton cross-section, which is a satisfactory description of the linear regime of QCD dynamics,  and the associated analytical expressions will be presented. Finally, in section \ref{conc}, our main conclusions are summarized and prospects are discussed. Four appendixes are also included, where some useful integrals are shown and the spinor matrix elements relevant for the calculations of the LCWFs are presented.



\section{The color - dipole $S$ - matrix framework and the  gauge boson production}
\label{sec:formalism_dijet}
At forward rapidities, the projectile parton has a large longitudinal momentum $x_1$ and the target structure is dominated by gluons, since $x_2 \ll 1$. As a consequence, the $pp$ cross-section is determined by the $a g \rightarrow b c$ subprocesses, where $b$ and $c$ can be both partons or a combination of a quark and an electroweak gauge boson for an incident quark ($a = q_f$). In the laboratory frame, such a process  can be viewed  as an excitation of the perturbative $|bc \rangle$ Fock state of the physical projectile $|a\rangle$ by a one - gluon exchange with the target proton \cite{Nikolaev:1994de,Nikolaev:1995ty}. At high energies, the parton $a$ can be assumed to propagate along a straight line with a fixed impact parameter. The perturbative transition $a \rightarrow bc$ can be described in terms of the Fock state expansion for the physical state $|a\rangle_{phys}$, which at the lowest order is given by \cite{Nikolaev:2003zf}
\begin{eqnarray}
|a\rangle_{phys} = |a\rangle_0 + \Psi(z_b,\rr)|bc\rangle_0 \,\,,
\end{eqnarray} 
where $|...\rangle_0$ refers to bare partons and $\Psi(z_b,\rr)$ is the probability amplitude to find the $bc$ system with separation $\rr$ in the two-dimensional impact parameter space. Considering that the impact parameter is conserved in the process, the action of the $S$ matrix on $|a\rangle_{phys}$ can be expressed as follows \cite{Nikolaev:2003zf}
\begin{eqnarray}
S|a\rangle_{phys} & = & S_a(\rb)|a\rangle_{0} + S_b(\rb_b)S_c(\rb_c)\Psi(z_b,\rr)|bc\rangle_0 \nonumber \\
& = & S_a(\rb)|a\rangle_{phys} + [S_b(\rb_b)S_c(\rb_c)-S_a(\rb)]\Psi(z_b,\rr)|bc\rangle \,\,,
\label{Eq:Smatrix}
\end{eqnarray}
where in the last line we have decomposed the final state into the elastically scattered $|a\rangle_{phys}$ and the excited state $|bc\rangle$. Moreover, one has assumed that the impact parameter of the parton $a$ is $\rb$, which implies $\rb_b = \rb + z_b\rr$ and $\rb_c = \rb - z_b\rr$, with $z_i$ the fraction of the longitudinal momentum of parton $a$ carried by the particle $i$. The last two terms in Eq. (\ref{Eq:Smatrix}) describe the scattering on the proton of the $bc$ system, which was formed before the interaction, and the transition $a \rightarrow bc$ after the interaction of $a$ with the proton. It is important to emphasize that the contribution associated with the transition $a \rightarrow bc$ inside the target vanishes in the high - energy limit. The differential cross-section for the  production of a dijet system, with momenta $\pv_b$ and $\pv_c$, is proportional to the modulus square of the scattering amplitude for the $a g \rightarrow b c$ process, which is given by \cite{Nikolaev:2004cu}
\begin{eqnarray}
{\cal{A}} = \int d^2\rb_b \, d^2\rb_c \exp[-i(\pv_b \cdot \rb_b + \pv_c \cdot \rb_c)[S_b(\rb_b)S_c(\rb_c)-S_a(\rb)]\Psi(z_b,\rr)\,\,. 
\end{eqnarray}
As a consequence, the master formula for the dijet production in the color - dipole $S$ - matrix framework is given by \cite{Nikolaev:2003zf}
\begin{eqnarray}
\label{Eq:master}
\frac{d\sigma (a \rightarrow b(p_b) c(p_c))}{dz d^2\pv_b d^2\pv_c} & = & \frac{1}{(2\pi)^4}\, 
\int d^2\bv_b d^2\bv_c d^2\bv_b^{\prime} d^2\bv_c^{\prime} \exp[i\pv_b \cdot (\bv_b - \bv_b^{\prime}) + i\pv_c \cdot (\bv_c - \bv_c^{\prime}) ]\, 
\Psi(z,\bv_b - \bv_c) \Psi^{*}(z,\bv_b^{\prime} - \bv_c^{\prime}) \nonumber \\
& \times & \left\{S^{(4)}_{\bar{b}\bar{c}cb}(\bv_b^{\prime},\bv_c^{\prime},\bv_b,\bv_c) + S^{(2)}_{\bar{a}a}(\bv^{\prime},\bv) - S^{(3)}_{\bar{b}\bar{c}a}(\bv,\bv_b^{\prime},\bv_c^{\prime}) - S^{(3)}_{\bar{a}bc}(\bv^{\prime},\bv_b,\bv_c) \right\}  \,\,,
\end{eqnarray}
where we have defined the quantities
\begin{eqnarray}
S_{a\bar{a}}^{(2)}(\rb^\prime,\rb) & = & S_a^\dagger(\rb^\prime)S_a(\rb) \,\,,\\
S_{\bar{a}bc}^{(3)}(\rb^\prime,\rb_b,\rb_c) & = & S_a^\dagger(\rb^\prime)S_b(\rb_b)S_c(\rb_c) \,\,, \\
S_{\bar{b}\bar{c}a}^{(3)}(\rb,\rb^\prime_b,\rb^\prime_c) & = & S_b^\dagger(\rb^\prime_b)S_c^\dagger(\rb^\prime_c)S_a(\rb) \,\,, \\
S_{\bar{b}\bar{c}cb}^{(4)}(\rb^\prime_b,\rb^\prime_c,\rb_b,\rb_c) & = & S_b^\dagger(\rb^\prime_b)S_c^\dagger(\rb^\prime_c)S_c(\rb_c)S_b(\rb_b)\,\,.
\end{eqnarray}
As the hermitian conjugate $S^\dagger$ can be viewed as the $S$ matrix for an antiparticle, one has that $S_{a\bar{a}}^{(2)}(\rb^\prime,\rb)$ represents the $S$ matrix for the interaction of the $a\bar{a}$ state with the target, with $\bar{a}$ propagating at the impact parameter $\rb^\prime$. The averaging over the color states of the beam parton $a$ implies that one has a color - singlet $a\bar{a}$ state. Similarly, $S_{\bar{a}bc}^{(3)}$ and $S_{\bar{b}\bar{c}cb}^{(4)}$ can be associated with the interaction of the color - singlet $\bar{a}bc$ and $\bar{b}\bar{c}cb$ systems, respectively. A detailed discussion about how to calculate these quantities in the general case was presented in a series of publications \cite{Nikolaev:2003zf,Nikolaev:2004cu,Nikolaev:2005dd,Nikolaev:2005zj,Nikolaev:2005ay,Nikolaev:2005qs}, which we refer for the interested reader. In what follows, we will focus on the $a \rightarrow G c$ process, with $G$ an electroweak gauge boson, which was not discussed in these previous studies.

For the production of an electroweak gauge boson, one has that the master equation simplifies, since they are color - singlet objects. In particular, for $b = G$ and $c = q$, we have the following simplifications:  $S_{\bar{b}\bar{c}cb}^{(4)} \rightarrow S_{q\bar{q}}^{(2)}$ and 
$S_{\bar{a}bc}^{(3)} \rightarrow S_{q\bar{q}}^{(2)}$. Therefore, the differential cross-section for the production of gauge boson associated with a quark is given by 
\begin{eqnarray}
\label{Eq:quark+G}
& \, & \frac{d\sigma^f_{T,L} (q_f p \rightarrow G(p_G) q_k(p_q))}{dz d^2\pv_G d^2\pv_q}  =  \frac{1}{(2\pi)^4}\, 
\int d^2\bv_G d^2\bv_q d^2\bv_G^{\prime} d^2\bv_q^{\prime} \exp[i\pv_G \cdot (\bv_G - \bv_G^{\prime}) + i\pv_q \cdot (\bv_q - \bv_q^{\prime}) ]\, \nonumber \\
& \times & \Psi_{T,L}(z,\bv_G - \bv_q) \Psi^{*}_{T,L}(z,\bv_G^{\prime} - \bv_q^{\prime}) 
 \left\{S^{(2)}_{q\bar{q}}(\bv_q^{\prime},\bv_q) + S^{(2)}_{q\bar{q}}(\bv^{\prime},\bv) - S^{(2)}_{q\bar{q}}(\bv,\bv_q^{\prime}) - S^{(2)}_{q\bar{q}}(\bv^{\prime},\bv_q) \right\} \,. 
\end{eqnarray}
Here, the $G$-boson and the final state quark share the light-cone plus momentum of the incoming quark in fractions $z$ and $1-z$,
respectively. Conservation of orbital angular momentum leads to the relations
\begin{eqnarray}
\bv = z \bv_G + (1-z) \bv_q  \, , \, \bv' = z \bv_G' + (1-z) \bv'_q \, ,
\end{eqnarray}
between the impact parameters of incoming and outgoing partons in amplitude and complex conjugate amplitude, respectively.
Introducing
\begin{eqnarray}
\rr = \bv_G - \bv_q \, , \, \rr' = \bv'_G - \bv'_q \, ,
\end{eqnarray}
we can thus express the relevant impact parameters as
\begin{eqnarray}
\bv_G &=& \bv + (1-z) \rr \, , \, \bv_q = \bv - z \rr \nonumber \\
\bv'_G &=& \bv' + (1-z) \rr' \, , \, \bv'_q = \bv' - z \rr' \, .
\end{eqnarray}
Changing variables in the impact parameter space integration, we can write
\begin{eqnarray}
\label{eq:Master_formula_reduced}
\frac{d\sigma^f_{T,L} (q_f \rightarrow G(p_G) q_k(p_q))}{dz d^2\pv_G d^2\pv_q} & = & \frac{1}{(2\pi)^4}\, 
\int  d^2\rr d^2\rr'  \exp[ i ((1-z)\pv_G- z \pv_q) \cdot (\rr - \rr^{\prime}) ] \Psi_{T,L}(z,\rr) \Psi^{*}_{T,L}(z,\rr') \nonumber \\
&\times&\int d^2\bv d^2\bv'\exp[i (\pv_G + \pv_q) \cdot (\bv - \bv^{\prime})] \nonumber \\
&\times&
 \left\{S^{(2)}_{q\bar{q}}(\bv'-z\rr',\bv-z\rr) + S^{(2)}_{q\bar{q}}(\bv^{\prime},\bv) - 
S^{(2)}_{q\bar{q}}(\bv'-z\rr',\bv) - S^{(2)}_{q\bar{q}}(\bv^{\prime},\bv-z\rr) \right\}  
\end{eqnarray}
One has that the conjugate variable to  $\rr - \rr'$ is the light-cone relative momentum
\begin{eqnarray}
\kv = (1-z) \pv_G - z \pv_q \, .
\end{eqnarray}
Introducing the transverse momentum decorrelation $\Deltav = \pv_G + \pv_q$, as well as
\begin{eqnarray}
\sv = \bv - \bv' \, , \, \Bv = {\bv + \bv' \over 2} \, ,
\end{eqnarray}
allows us to express the differential cross-section in terms of the $q \bar q$ dipole cross-section after integrating out $\Bv$.
For brevity of notation, we change to the cross-section differential in the variables $\kv$ and $\Deltav$, which is given by
\begin{eqnarray}
\label{Eq:Master_dijet}
\frac{d\sigma^f_{T,L} (q_f \rightarrow G(p_G) q_k(p_q))}{dz d^2\kv d^2\Deltav} & = & \frac{1}{2 (2\pi)^4}\, 
\int  d^2\rr d^2\rr'  \exp[ i \kv \cdot (\rr - \rr^{\prime}) ] \Psi_{T,L}(z,\rr) \Psi^{*}_{T,L}(z,\rr') \nonumber \\
&\times&\int d^2\sv \exp[i \Deltav \cdot \sv] 
 \Big\{
 \sigma_{q\bar{q}}(\sv+z\rr') + \sigma_{q\bar{q}}(\sv-z\rr) 
- \sigma_{q\bar{q}}(\sv -z(\rr-\rr') ) - \sigma_{q\bar{q}}(\sv)
\Big\}  \,\,. 
\end{eqnarray}
Here we made use of the definition of the dipole cross-section in terms of the $q \bar q$ $S$-matrix:
\begin{eqnarray}
\sigma(\rr) = 2 \int d^2\Bv \, \Big[ 1 - S^{(2)}_{q \bar q} \Big(\Bv + {\rr \over 2}, \Bv - {\rr \over 2} \Big)\Big] \, .
\end{eqnarray}
Equation (\ref{Eq:Master_dijet}) is the main ingredient to estimate the associated production of an electroweak gauge boson with a hadron, which will be discussed in a forthcoming publication. Here, we focus on the isolated gauge boson production, which has a cross - section that can be derived  integrating the master equation over the transverse momentum of the quark in the final state or, equivalently,  integrating  Eq. (\ref{Eq:Master_dijet})  over $\Deltav$.

Assuming that the projectile quark is unpolarized and integrating  Eq. (\ref{Eq:Master_dijet})  over $\Deltav$,  one has that the differential cross-section for the $q_f p \rightarrow GX$ process reads
\begin{eqnarray}
\label{Eq:isolated}
\frac{d\sigma^f_{T,L} (q_fN \rightarrow GX)}{dz d^2\pv} & = & \frac{1}{(2\pi)^2}\, \overline{\sum}_\text{quark pol.}
\int d^2\rr \dd^2\rp \exp[i\pv \cdot (\rr - \rp)]\, \Psi_{T,L}(z,\rr) 
\Psi^{*}_{T,L}(z,\rp) \nonumber \\
 & \times & \frac{1}{2}\left[ \sigma_{q\bar{q}}(z\rr,x) + \sigma_{q\bar{q}}(z\rp,x) - 
 \sigma_{q\bar{q}}(z|\rr - \rp|,x)\right]\,, 
\end{eqnarray}
where  $\rr$ and $\rp$ are the quark-$G$ transverse separations in the total radiation 
amplitude and its conjugated, respectively.  Moreover, the average over quark polarization implies that 
\begin{eqnarray} \label{Psi2}
&  & \overline{\sum}_\text{quark pol.} \Psi_{T,L}(z,\rr,m_f) \Psi^{*}_{T,L}(\,\rp,m_f)  =  \rho^V_{T,L}(z,\rr,m_f) + \rho^A_{T,L}(z,\rr,m_f) \,,
\end{eqnarray}
with 
\begin{eqnarray}\label{Psi2b}
    \rho^T_V(z,\rr,\rr') &=& \frac{1}{2} \sum_{\lambda, \lambda', \lambda_G} \Psi^{T,\lambda_G}_V(z,\rr) \Psi^{T, \lambda_G *}_V(z,\rr'),  \qquad
      \rho^T_A(z,\rr,\rr') = \frac{1}{2} \sum_{\lambda, \lambda', \lambda_G} \Psi^{T,\lambda_G}_A(z,\rr) \Psi^{T, \lambda_G *}_A(z,\rr')  \, , \nonumber \\
        \rho^L_V(z,\rr,\rr') &=& \frac{1}{2} \sum_{\lambda, \lambda'} \Psi^{L}_V(z,\rr) \Psi^{L*}_V(z,\rr'),  \qquad 
      \rho^L_A(z,\rr,\rr') = \frac{1}{2} \sum_{\lambda, \lambda'} \Psi^{L}_A(z,\rr) \Psi^{L*}_A(z,\rr')  \, .
\end{eqnarray}
Therefore, the transverse momentum spectrum for the electroweak gauge boson production in the $q p \rightarrow GX$ channel is fully determined by the dipole - proton cross - section $\sigma_{q\bar{q}}$, which is associated with the QCD dynamics, and by the square of the light cone wave functions, represented by the functions $\rho_i^j$, defined in Eq. (\ref{Psi2b}). In the next section, we will present a detailed derivation from these quantities.

\section{Light cone wave function of the gauge boson radiation off a quark}
\label{sec:wf}
\begin{figure}[t]
{\includegraphics[width=0.5\textwidth]{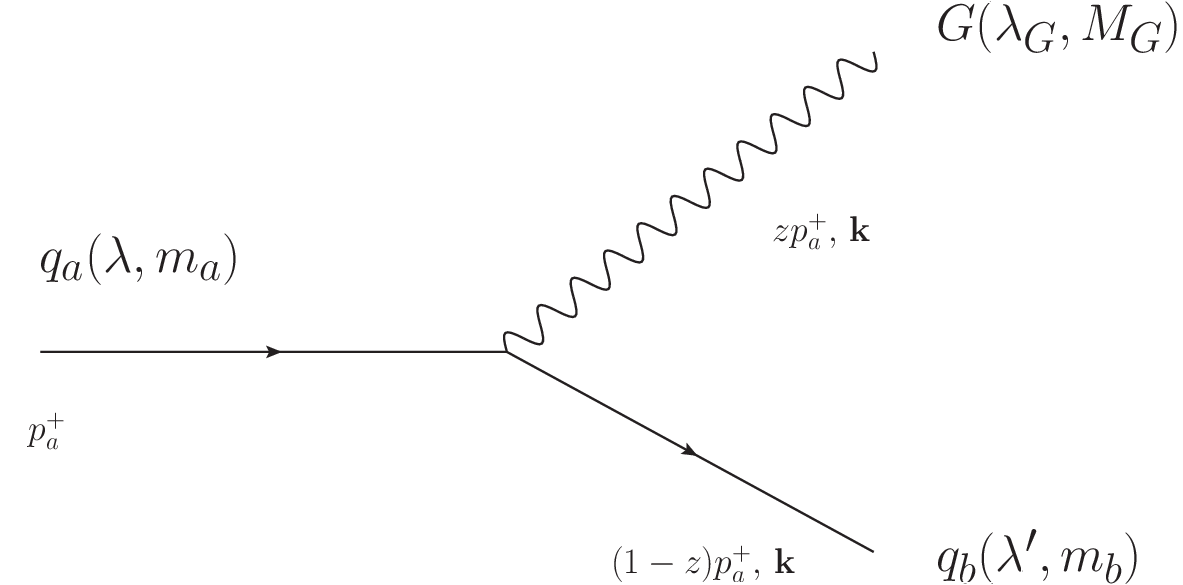}}  
\caption{Diagram representing the emission of a gauge boson of mass $M_G$ and polarization $\lambda_G$ by an incident quark of mass $m_a$ and polarization $\lambda$ that becomes a quark of mass $m_b$ and polarization $\lambda^{\prime}$. For  $\gamma$ and $Z^0$ radiation one
has $q_a = q_b$, which implies $m_a =  m_b$.}
\label{fig:diagram2}
\end{figure}   
\begin{table}[t]
\begin{center}

\begin{tabular}{||c|c|c|c||}\hline \hline
     Gauge Boson               & ${\cal C}^G_f$ & $g_{v,f}^G$  & $g_{a,f}^G$   \\ \hline \hline
$Z^0$ & $ {\cal C}^Z_f=\frac{\sqrt{\alpha_{em}}}{\sin 2\theta_W}$ & $g_{v,f_u}^Z=\frac12-\frac43\sin^2\theta_W$ & $g_{a,f_u}^Z=\frac12$ \\ 
   &        & $g_{v,f_d}^Z=-\frac12+\frac23\sin^2\theta_W $ & $g_{a,f_d}^Z=-\frac12$ \\ \hline 
   $W^{\pm}$ &  $ {\cal C}^{W^+}_f=\frac{\sqrt{\alpha_{em}}}{2 \sqrt{2}\sin \theta_W}V_{f_uf_d}$       &  $g_{v,f}^{W}$ = 1      &  $g_{a,f}^{W}$ = 1        \\
       &  $ {\cal C}^{W^-}_f=\frac{\sqrt{\alpha_{em}}}{2 \sqrt{2}\sin \theta_W}V_{f_df_u}$       &        &         \\\hline 
   Photon    & $C_f^{\gamma} = \sqrt{\alpha_{em}} e_f$       & $g_{v,f}^{\gamma}$ = 1       &  $g_{a,f}^{\gamma}$ = 0       \\ \hline \hline 
\end{tabular} 
\caption{Vector and axial couplings and coefficients for the distinct gauge bosons, with $\theta_W$ being the Weinberg angle, $f_u=u,c,t$ and $f_d=d,s,b$ are the flavors of up- and down-type quarks, respectively, and $V_{f_uf_d}$ the corresponding CKM matrix elements.  }
\label{Tab:factors}
\end{center}
\end{table} 
We now turn to the evaluation of the LCWF for the $q \to G q'$ transition. Here we wish to choose light front(LF)-gauge for all the gauge boson fields, i.e. 
\begin{eqnarray}
n^-_\mu A^\mu = n^-_\mu W^{\mu \pm} = n^-_\mu Z^\mu = 0.    
\end{eqnarray}
In this gauge, the LCWF  can be calculated from the matrix element of the appropriate part of the LF-Hamiltonian. This is standard procedure in the case of QED, but becomes a bit more involved in the case of massive gauge bosons. Here one needs to start from the full Lagrangian of the Standard-Model, including the would-be Goldstone boson fields of the broken gauge symmetry. Apparently, this approach to LCWFs of gauge bosons has not been previously presented in the literature. We therefore start from reminding the reader of some details of the Standard Model Lagrangian. Here we follow closely the presentation in \cite{Srivastava:2002mw}.

Let us discuss the relevant vertices on the example of the first generation of quarks $(u,d)$.
With respect to the Standard Model gauge group $SU(2)_L \times U(1)_Y$, we group the left-handed fermions into a doublet, while right-handed fermions are singlets. To lighten up the notation, we omit the mixing of down-type flavours by the  CKM matrix, which we will restore in our final results for the $W^\pm$ bosons. Therefore, we will assume
\begin{eqnarray}
\psi_L = \begin{pmatrix}
    u \\ d
\end{pmatrix}_L \, , \quad  u_R, d_R \, .
\end{eqnarray}
Then, for massless quarks coupling to electroweak bosons is obtained from the Lagrangian
\begin{eqnarray}
    {\cal L} = \bar \psi_L i \gamma^\mu D_\mu \psi_L + \bar u_R i \gamma^\mu D_\mu u_R + \bar d_R i \gamma^\mu D_\mu d_R
    \label{eq:Lagrangian}
\end{eqnarray}
with the covariant derivative
\begin{eqnarray}
    D_\mu = \partial_\mu + \frac{g}{\sqrt{2}}\Big( W^+_\mu (t_1 + i t_2) + W^-_\mu (t_1 - i t_2) \Big) - \frac{g}{\cos \theta_W} Z_\mu (t_3 - Q \sin^2 \theta_W) - i e Q A_\mu   
\end{eqnarray}
The generators $\vec t = (t_1,t_2,t_3)$ for the $SU(2)_L$ doublet are $\vec t = \vec \tau /2$, with $\tau_1, \tau_2,\tau_3$ being the Pauli matrices, while acting on the singlet fields $\vec t \equiv 0$.
The electric charge is $Q = t_3 + Y$, with $Y$ being the weak hypercharge.
Then, from Eq. (\ref{eq:Lagrangian}) we obtain the interaction terms
\begin{eqnarray}
e A_\mu J^\mu_{\rm em} + g\Big( W^+_\mu J_W^{\mu} + W^-_\mu J_W^{\dagger \mu} + Z_\mu J_Z^\mu \Big) \, ,      \end{eqnarray}
where 
\begin{eqnarray}
J^\mu_{\rm em} &=& Q_u \, \bar u \gamma^\mu u + Q_d \,  \bar d \gamma^\mu d \nonumber \\
J_W^{\mu} &=& \frac{1}{2\sqrt{2}} \bar u \gamma^\mu (1-\gamma_5) d \, , \quad 
J_W^{\dagger \mu} = \frac{1}{2 \sqrt{2}} \bar d \gamma^\mu (1- \gamma_5) u  \nonumber \\
J_Z^\mu  &=& \frac{1}{\cos \theta_W}\Big( \frac{1}{4} \bar u \gamma^\mu (1-\gamma_5) u - \frac{1}{4} \bar d \gamma^\mu (1-\gamma_5) d - \sin^2 \theta_W J^\mu_{\rm em} \Big)  \, . 
\end{eqnarray}
Besides the interaction vertices deriving from this part of the Standard Model Lagrangian, we also need the contributions from the gauge-invariant Yukawa interactions:
\begin{eqnarray}
{\cal L}_{\rm Yuk} =    - y_d (\bar \psi_L \Phi) d_R  - y_u (\bar \psi_L i\tau_2 \Phi^*) u_R  + {\rm h.c.} \, .
   \label{eq:Yukawa}
\end{eqnarray}
The scalar field is in a doublet representation of $SU(2)_L$ and parametrized as
\begin{eqnarray}
    \Phi = \begin{pmatrix} \phi^+ \\ \frac{1}{\sqrt{2}} ( v + h + i \phi^0 ) 
    \end{pmatrix} \, ,
\end{eqnarray}
with the vacuum expectation value
\begin{eqnarray}
    \bra{0} | \Phi | \ket{0} =\frac{1}{\sqrt{2}} \begin{pmatrix} 0 \\ v
    \end{pmatrix} \, ,
\end{eqnarray}
the Higgs field $h$ and the would-be Goldstone boson $\phi$. The Yukawa interactions given by Eq. (\ref{eq:Yukawa}) lead to the quark masses $m_u = y_u v/\sqrt{2}, m_d = y_d v/\sqrt{2}$. In the gauge sector, we have $M_W = gv/2, M_Z = M_W/\cos \theta_W$.
As we wish to use the LF gauge, where 
\begin{equation}
n^-_\mu  A^\mu  = n^-_\mu Z^\mu = n^-_\mu W^{\mu \pm} = 0, 
\end{equation}
we cannot ``gauge away'' the Goldstone field, but have to take into account the vertices with the field $\phi$, which are, omitting fermion mass-terms and Higgs-couplings:
\begin{eqnarray}
{\cal L}_{\rm{Yuk}} &\supset& \Big( y_u \bar u_R d_L - y_d \bar u_L d_R \Big) \phi^+ 
+ \Big( y_u \bar d_L u_R - y_d \bar d_R u_L \Big) \phi^- + \frac{i}{\sqrt{2}} \Big( y_u \bar u \gamma_5 u - y_d \bar d \gamma_5 d\Big) \phi^0 \nonumber \\ 
&=& \frac{g}{2 \sqrt{2} M_W^2} \Big( (m_u - m_d) \bar u d - (m_u + m_d) \bar u \gamma_5 d \Big) (-i \partial_\mu W^{\mu+}) \nonumber \\
&+& \frac{g}{2 \sqrt{2} M_W^2} \Big( (m_u - m_d) \bar d u + (m_u + m_d) \bar d \gamma_5 u \Big) (i \partial_\mu W^{\mu-}) \nonumber \\
&+& \frac{ig}{2 \cos \theta_W M_Z^2} \Big( m_u \bar u \gamma_5 u - m_d \bar d \gamma_5 d \Big) \partial_\mu Z^\mu \, . 
\end{eqnarray}
Here we have anticipated, that free fields satisfy the 't Hooft conditions
\begin{eqnarray}
    M_W \phi^\pm = \pm i \partial_\mu W^{\mu \pm} \, , \, M_Z \phi^0 = \partial_\mu Z^\mu \, . 
\end{eqnarray}
Now we have collected all the ingredients for the calculation of the LCWFs.
Let us start from the $W^- u$ Fock state component of the $d$-quark.
\begin{eqnarray}
    \delta(p_b^+ + p_c^+ - p_a^+) \delta^{(2)}(\pv_b + \pv_c)  \psi_{W^-u/d}(z,\pv_b) = \, \frac{\Bigbra{W^-(p_c^+,\pv_c,\lambda_W) u(p^+_b, \pv_b,\lambda')} \Big| {\cal H}_{\rm int}^{\rm LF}\Big|\Bigket{d(p_a^+,{\bf 0},\lambda)}}{p_a^- - p_b^- - p_c^-} \, .  \nonumber \\
\end{eqnarray}
For the numerator, we need to evaluate
\begin{eqnarray}
    E^*_\mu(k,\lambda_W) \bar u_u(p_b,\lambda') \Big\{ \gamma^\mu (1-\gamma_5) + \frac{k^\mu}{M^2_W} \Big( (m_u - m_d) - (m_u + m_d) \gamma_5 \Big) \Big \} u_d(p_a,\lambda) \, . 
    \label{eq:Numerator}
\end{eqnarray}
Here, the contribution$\propto \gamma^\mu (1-\gamma_5)$ derives from the gauge-boson part of the Lagrangian, while the part $\propto k^\mu$ is the Goldstone boson contribution.
The spinors for $u$ and $d$ quarks fulfill the respective Dirac equations,
\begin{eqnarray}
    (\gamma \cdot p_a) u_d(p_a,\lambda) = m_d\, , \quad (\gamma \cdot p_b) u_u(p_b,\lambda') = m_u \, .
    \label{eq:Dirac-eqn}
\end{eqnarray}
The polarization vectors for transversely polarized ($\lambda_W = \pm 1$) are the standard ones that one would also use for photons:
\begin{eqnarray}
    E_\mu(k,\pm 1) = E_\mu^{\perp}(\pm 1) - \frac{E^{\perp}(\pm 1)\cdot k}{k^+} n^-_\mu \, . 
\end{eqnarray}
The transverse polarization vectors fulfill 
\begin{eqnarray}
n_\mu^- E^\mu(k,\pm 1) = 0, \quad k_\mu E^\mu(k,\pm 1) = 0 \, , 
\end{eqnarray}
where the latter property ensures, that the would-be Goldstones do not contribute for transversely polarized gauge bosons.

The situation is a bit more subtle for the case of longitudinal polarizations. In LF-gauge, the third polarization vector of the gauge boson field has the form \cite{Srivastava:2002mw}:
\begin{eqnarray}
    E_\mu(k,0) = - \frac{M_W}{k^+} \, n^-_\mu \, . 
\end{eqnarray}
It does fulfill the LF-gauge condition $n^-_\mu E^\mu(k,0) =0$, but is not orthogonal to $k_\mu$, instead we have
\begin{eqnarray}
    k_\mu E^\mu(k,0) = - M_W \, . 
\end{eqnarray}
Let us now turn to the evaluation of the LF-wave function.
Firstly, the four momenta of the on-shell particles read in component form:
\begin{eqnarray}
p_a = \Big[p_a^+, \frac{m_d^2}{2 p_a^+},{\bf 0 }\Big] \, , \, k \equiv p_c =\Big[ z p_a^+, \frac{\kv^2 + M_W^2}{2 z p_a^+}, \kv\Big] \, , \, p_b = \Big[(1-z)p_a^+, \frac{\kv^2 + m_u^2}{2(1-z)p_a^+},-\kv\Big] \, ,
\end{eqnarray}
so that we obtain for the LF-``energy denominator''
\begin{eqnarray}
    p_a^- - p_b^- - k^- = -\frac{\kv^2 + \epsilon^2}{2 z(1-z) p_a^+}  \, , 
\end{eqnarray}
with 
\begin{eqnarray}
\epsilon^2 = (1-z) M_W^2 + z m_u^2 - z(1-z) m_d^2 = (1-z) M_W^2 + z (m_u^2 - m_d^2) + z^2 m_d^2 \, .
\end{eqnarray}
For transverse polarization, we obtain then (we omit the helicity labels)
\begin{eqnarray}
\psi^T_{W^- u/d}(z,\kv) = -\frac{ z(1-z) g}{2 \sqrt{2}}  
\frac{ E^*_\mu(\pm 1,k) \bar u_u(p_b,\lambda')  \gamma^\mu (1-\gamma_5) u_d(p_a,\lambda)}{\kv^2 + \epsilon^2} \, , 
\end{eqnarray}
while for longitudinally polarized $W$'s we have
\begin{eqnarray}
\psi^L_{W^-u/d}(z,\kv) = \frac{ z(1-z)g}{2 \sqrt{2}M_W} \frac{ \bar u_u(p_b,\lambda') \Big\{ \frac{M^2_W}{k^+} \gamma^+ (1-\gamma_5) +  (m_u - m_d) - (m_u + m_d) \gamma_5  \Big \} u_d(p_a,\lambda) }
{\kv^2 + \epsilon^2} \nonumber \\
\end{eqnarray}
Explicit results can be obtained by utilizing the spinor matrix elements given in  Appendix \ref{ap:spinor}.

It is interesting to compare the results for the ones obtained in the literature in unitary gauge.
The polarization vector for the third, longitudinal polarization of an on-shell massive spin-one particle reads
\begin{eqnarray}
    \tilde E_\mu(k,0) = \frac{k_\mu}{M_W} - \frac{M_W}{k^+} n^-_\mu = \frac{k_\mu}{M_W} + E_\mu(k,0) \, , 
\end{eqnarray}
and fulfills
\begin{eqnarray}
    k^\mu \tilde E_\mu(k,0) = 0 \, , \quad \tilde E_\mu(k,0) \tilde E^\mu(k,0) = -1 \, . 
\end{eqnarray}
Then, let us use the polarization vector $\tilde E_\mu(k,0)$ to evaluate the matrix element of Eq.(\ref{eq:Numerator}). Evidently, the piece from the would-be Goldstone bosons now cancels, but from the gauge boson part we obtain the additional contribution
\begin{eqnarray}
    \bar u_u(p_b, \lambda') (k\cdot \gamma)(1-\gamma_5) u_d(p_a,\lambda) \, . 
\end{eqnarray}
If we want to use the equations of motion for quarks, Eq.(\ref{eq:Dirac-eqn}), we need to be careful, because four-momentum is not conserved in our calculation of the LCWF. Instead:
\begin{eqnarray}
 k_\mu = p_{a\mu} - p_{b\mu} + (k^- + p_b^- - p_a^-) \, n^-_\mu \, . 
\end{eqnarray}
Here the first two terms will indeed reproduce the Goldstone boson contribution of the LF-gauge, while the third term will cancel the energy denominator of the LCWF and will give an additive contribution to the WF that does not depend on $\kv$.
This constant piece in impact parameter space gives rise to a contribution to the WF $\propto \delta^{(2)}(\rr)$.  It therefore will drop out of all observables of interest in this work. However, the result of the ``unitary gauge'' calculation clouds up the interpretation of the square of the LFWF as a probability of a two-body Fock state in the quark. The origin of this trouble is of course 
the $\propto k_\mu/M_W$ piece in the polarization vector of the massive vector, which is taken care of by the would-be Goldstone fields in LF-gauge.

Finally, to wrap up our results in a way comprising all cases $G = W^\pm, Z^0, \gamma^*$, we introduce the LCWFs of an effective ``vector'' and ``axial vector'' boson as in the $q_q \to q_b G$ transition. We absorb a factor $1/\sqrt{z(1-z)}$ into the WF so that the phase space is $\propto dz$. 
\begin{eqnarray}
\Psi_{V}(z,\kv) =  \, C_f^G g_{V,f}^G \, \sqrt{z(1-z)} \, \frac{\Gamma_V}{\kv^2 + \epsilon^2} \, , \, \Psi_{A}(z,\kv) =  \, C_f^G g_{A,f}^G \,  \sqrt{z(1-z)} \, \frac{\Gamma_A} {\kv^2 + \epsilon^2} \, .
\end{eqnarray}
with 
\begin{eqnarray}
    \Gamma_V &=&     E^*_\mu(k,\lambda_W) \bar u(p_b,\lambda',m_b) \Big\{ \gamma^\mu  +  (m_b - m_a) \frac{k^\mu}{M^2_G} \Big \} u(p_a,\lambda,m_a) \, \nonumber \\
    \Gamma_A &=&    E^*_\mu(k,\lambda_W) \bar u(p_b,\lambda',m_b) \Big\{ \Big(\gamma^\mu  + (m_a + m_b)\frac{k^\mu}{M^2_G} \Big) \gamma_5   \Big \} u(p_a,\lambda,m_a) \, .
\end{eqnarray}
The relevant couplings $C_f^G, g^G_{V,f}, g^G_{A,f}$ are collected in Table \ref{Tab:factors}.
Let us collect explicit expressions for the LCWFs. Firstly, for transverse polarizations $\lambda_G = \pm 1$, we have
\begin{eqnarray}
\Psi^{T, \lambda_G}_V(z,\kv) &=&  C_f^G g_{V,f}^G \sqrt{z(1-z)} \, 
\chi^\dagger_{\lambda'} \Big(- \bj\cdot \bE^*(\lambda_G) + \frac{\kv \cdot \bE^*(\lambda_G)}{z p_a^+}  j^+ \Big) \chi_\lambda  \, \, \frac{1}{\kv^2 + \epsilon^2} \, \nonumber \\
&=&  C_f^G g_{V,f}^G \frac{\sqrt{z}}{\kv^2 + \epsilon^2} \, \chi^\dagger_{\lambda'} \Big\{ \Big( \frac{2 - z}{z}  \kv\cdot \bE^*(\lambda_G) + i \lambda [\kv,\bE^*(\lambda_G)]\Big) \openone - (m_b - (1-z)m_a) \, \lambda \bsigma\cdot \bE^*(\lambda_G) \Big\} \chi_\lambda \, ,
\nonumber \\
\end{eqnarray}
where the polarization state of the quark is given by the Pauli-spinor $\chi_\lambda$.
For longitudinal gauge boson polarizations, we have
\begin{eqnarray}
\Psi^L_V(z,\kv) &=&
 C_f^G g_{V,f}^G \frac{1}{\sqrt{z}} \frac{1}{M_G}
\chi^\dagger_{\lambda'} \Big\{ 
\Big( \frac{z^2m_a(m_b-m_a) -z (m_b^2 - m_a^2) - 2 (1-z) M_G^2}{\kv^2 + \epsilon^2} \Big) \openone + z(m_b - m_a) \frac{\lambda \, \bsigma \cdot \kv}{\kv^2 + \epsilon^2} \Big\} \chi_\lambda \nonumber \\
\end{eqnarray}
For the axial vector coupling, we obtain for the transverse polarizations
\begin{eqnarray}
\Psi^{T, \lambda_G}_A(z,\kv) =  C_f^G g_{A,f}^G \frac{\sqrt{z}}{\kv^2 + \epsilon^2} \, \chi^\dagger_{\lambda'} \Big\{ \Big( \frac{2 - z}{z}  \kv\cdot \bE^*(\lambda_G) + i \lambda [\kv,\bE^*(\lambda_G)]\Big) \lambda \openone - (m_b + (1-z)m_a) \,  \bsigma\cdot \bE^*(\lambda_G) \Big\} \chi_\lambda \nonumber \\
\end{eqnarray}
and for the longitudinal case we have
\begin{eqnarray}
\Psi^L_A(z,\kv) &=&  C_f^G g_{A,f}^G \frac{1}{\sqrt{z}} \frac{1}{M_G} \chi^\dagger_{\lambda'} \Big\{ \Big( - \frac{z^2m_a(m_b+m_a) +z (m_b^2 - m_a^2) + 2 (1-z) M_G^2}{\kv^2 + \epsilon^2} \Big) \lambda \openone - z(m_b + m_a) \frac{\bsigma \cdot \kv}{\kv^2 + \epsilon^2} \Big\} \chi_\lambda \nonumber \\
\end{eqnarray}
We now obtain the LCWFs in the mixed $z,\rr$ representation
\begin{eqnarray}
\Psi(z,\rr) = \int \frac{d^2 \kv}{(2 \pi)^2} \, \exp(-i \kv \cdot \rr) \, \Psi(z,\kv) \, .
\end{eqnarray}
For the transverse wave functions, we obtain
\begin{eqnarray}
 \Psi^{T,\lambda_G}_V(z,\rr) &=&  C_f^G g_{V,f}^G \frac{1}{2 \pi} \frac{1}{\sqrt{z}} \, \chi^\dagger_{\lambda'} 
 \Big\{ \Big( (2-z) \, \frac{\rr \cdot\bE^*(\lambda_g)}{r} + z i \lambda   \frac{[\rr,\bE^*(\lambda_g)]}{r} \Big) \openone \, i \epsilon K_1(\epsilon r)\nonumber \\
 &&- \lambda \bsigma\cdot \bE^*(\lambda_G) \, (z(m_b-m_a) + z^2 m_a) \, K_0(\epsilon r) 
 \Big\} \chi_\lambda\, \nonumber \\
\Psi^{T,\lambda_G}_A(z,\rr) &=&  C_f^G g_{A,f}^G \frac{1}{2 \pi} \frac{1}{\sqrt{z}} \, \chi^\dagger_{\lambda'} 
 \Big\{ \Big( (2-z) \, \frac{\rr \cdot\bE^*(\lambda_g)}{r} + z i \lambda   \frac{[\rr,\bE^*(\lambda_g)]}{r} \Big) \, \lambda \openone \,  i \epsilon K_1(\epsilon r)\nonumber \\
 &&-  \bsigma\cdot \bE^*(\lambda_G) \, (z(m_b+m_a) - z^2 m_a) \, K_0(\epsilon r) 
 \Big\} \chi_\lambda\, \, ,
\end{eqnarray}
while the longitudinal functions become
\begin{eqnarray}
    \Psi^L_V(z,\rr) &=&  C_f^G g_{V,f}^G \frac{1}{2 \pi} \frac{1}{\sqrt{z}} \frac{1}{M_G} \, 
    \chi^\dagger_{\lambda'} \Big\{ \Big( z^2 m_a(m_b - m_a) - z (m_b^2 - m_a^2) - 2(1-z)M_G^2 \Big) \, \openone \,  K_0(\epsilon r) \nonumber \\
    &&+ z(m_b - m_a) \, \lambda \frac{\bsigma \cdot \rr}{r} \, i \epsilon K_1(\epsilon r) \Big\} \chi_\lambda \nonumber \\
    \Psi^L_A(z,\rr) &=&  C_f^G g_{A,f}^G \frac{1}{2 \pi} \frac{1}{\sqrt{z}} \frac{1}{M_G} \, 
    \chi^\dagger_{\lambda'} \Big\{ - \Big( z^2 m_a(m_b + m_a) + z (m_b^2 - m_a^2) + 2(1-z)M_G^2 \Big) \, \lambda \, \openone \,  K_0(\epsilon r) \nonumber \\
    &&- z(m_b + m_a) \, \frac{\bsigma \cdot \rr}{r} \, i \epsilon K_1(\epsilon r) \Big\} \chi_\lambda 
\end{eqnarray}
Calculating the combinations
\begin{eqnarray}
    \rho^T_V(z,\rr,\rr') &=& \frac{1}{2} \sum_{\lambda, \lambda', \lambda_G} \Psi^{T,\lambda_G}_V(z,\rr) \Psi^{T, \lambda_G *}_V(z,\rr'),  \qquad
      \rho^T_A(z,\rr,\rr') = \frac{1}{2} \sum_{\lambda, \lambda', \lambda_G} \Psi^{T,\lambda_G}_A(z,\rr) \Psi^{T, \lambda_G *}_A(z,\rr')  \, , \nonumber \\
        \rho^L_V(z,\rr,\rr') &=& \frac{1}{2} \sum_{\lambda, \lambda'} \Psi^{L}_V(z,\rr) \Psi^{L*}_V(z,\rr'),  \qquad 
      \rho^L_A(z,\rr,\rr') = \frac{1}{2} \sum_{\lambda, \lambda'} \Psi^{L}_A(z,\rr) \Psi^{L*}_A(z,\rr')  \, .
\end{eqnarray}
one has that
\begin{eqnarray}
\rho^T_V(z,\rr,\rr') &=& \frac{({\cal C}^G_f)^2(g^{G}_{V,f})^2}{2\pi^2}  \Big\{ \frac{1 + (1-z)^2}{z} \, \frac{\rr \cdot \rr'}{rr'} \epsilon^2 K_1(\epsilon r) K_1(\epsilon r') + z\Big((m_b - m_a) + z m_a\Big)^2 K_0(\epsilon r) K_0(\epsilon r') \Big\} \nonumber \\  
\rho^T_A(z,\rr,\rr') &=& \frac{({\cal C}^G_f)^2(g^{G}_{A,f})^2}{2\pi^2}  \Big\{ \frac{1 + (1-z)^2}{z} \, \frac{\rr \cdot \rr'}{rr'} \epsilon^2 K_1(\epsilon r) K_1(\epsilon r') + z\Big((m_b + m_a) - z m_a\Big)^2 K_0(\epsilon r) K_0(\epsilon r') \Big\} \nonumber \\
\rho^L_V(z,\rr,\rr') &=& \frac{({\cal C}^G_f)^2(g^{G}_{V,f})^2}{4\pi^2} 
\Big\{ \frac{(z^2 m_a(m_b - m_a) - z (m_b^2 - m_a^2) - 2 (1-z) M_G^2)^2}{z M_G^2} \, K_0(\epsilon r) K_0(\epsilon r') \nonumber \\
&& + \frac{z (m_b - m_a)^2}{M_G^2}  \, \frac{\rr \cdot \rr'}{rr'} \, \epsilon^2 K_1(\epsilon r) K_1(\epsilon r') \Big\} \nonumber \\
\rho^L_A(z,\rr,\rr') &=& \frac{({\cal C}^G_f)^2(g^{G}_{A,f})^2}{4\pi^2}  
\Big\{ \frac{(z^2 m_a(m_b + m_a) + z (m_b^2 - m_a^2) + 2 (1-z) M_G^2)^2}{z M_G^2} \, K_0(\epsilon r) K_0(\epsilon r') \nonumber \\
&& + \frac{z (m_b + m_a)^2}{M_G^2} \, \frac{\rr \cdot \rr'}{rr'} \, \epsilon^2 K_1(\epsilon r) K_1(\epsilon r') \Big\} 
\end{eqnarray}
In the next section, we will use these results in Eq. (\ref{Eq:isolated}) to derive the expressions for the spectrum associated with the isolated gauge boson production. Such expressions reduce to  those derived in Refs.  \cite{kst99,Pasechnik:2012ac}  for $G = \gamma^*$ and $Z^*$ using different frameworks.

\section{Gauge boson production in the transverse momentum and impact parameter representations}
\label{sec:representations}

As demonstrated in section \ref{sec:formalism_dijet}, the differential cross - section for the isolated gauge boson production is expressed in terms of the squared LCWF's, derived in the previous section,  and the dipole - proton cross - section $\sigma_{q\bar{q}}$. As this quantity is usually derived in the impact parameter space, e.g., by solving the Balitsky - Kovchegov equation \cite{BAL,KOVCHEGOV}, a natural representation of the spectrum is at this same space. The associated expressions are derived in the subsection \ref{subsec:imp}. However, it is also possible to use the relation between the dipole - proton cross-section and the unintegrated gluon distribution $f(x,\kv)$ given by
\begin{eqnarray}
\sigma(\rr ,x) =\frac{1}{2} \int d^2 \kv f(x,\kv) \, \Big( 2 - e^{i \kv \cdot \rr}   - e^{-i \kv \cdot \rr} \Big) \,\,,
\label{Eq:unint}
\end{eqnarray}
and represent the spectrum in the transverse momentum space. In subsection 
\ref{subsec:mom} we derive the corresponding expressions for the spectrum in this representation. It is important to emphasize that both representations are equivalent, and approximated limits of both have been used in the literature to estimate the Drell - Yan process and the real photon production. The choice of one particular representation is, in general, associated with a description of the QCD dynamics, which sometimes is performed in terms of $\sigma_{q\bar{q}}$ and in other cases of $f(x,\kv)$.

\subsection{Representation in the impact momentum space}
\label{subsec:imp}
In what follows, we will derive in detail the contribution for a transverse polarization. One has that 
\begin{eqnarray}
\frac{d\sigma^f_{T}}{dz d^2\pv} & = & \frac{1}{2(2\pi)^2}\,\int d^2\rr \dd^2\rp e^{i\pv \cdot (\rr - \rp)}
\rho^T_V(z,\rr,\rr') \left[ \sigma_{q\bar{q}}(z\rr,x) + \sigma_{q\bar{q}}(z\rp,x) - 
 \sigma_{q\bar{q}}(z|\rr - \rp|,x)\right] \nonumber \\
& + & \frac{1}{2(2\pi)^2}\,\int d^2\rr \dd^2\rp e^{i\pv \cdot (\rr - \rp)}
\rho^T_A(z,\rr,\rr') \left[ \sigma_{q\bar{q}}(z\rr,x) + \sigma_{q\bar{q}}(z\rp,x) - 
 \sigma_{q\bar{q}}(z|\rr - \rp|,x)\right]
\end{eqnarray}
Substituting the expression for $\rho^T_V(z,\rr,\rr')$,     one has that the vector contribution can be expressed by:
\begin{eqnarray}
z\frac{d\sigma^f_{T}}{dz d^2\pv}|_V & = & \frac{1}{2(2\pi)^2}\, \frac{({\cal C}^G_f)^2(g^{G}_{v,f})^2}{2\pi^2} z^2[(m_b - m_a) + zm_a]^2 \, \nonumber \\
& \times & \,\int d^2\rr \dd^2\rp e^{i\pv \cdot (\rr - \rp)}
      {\rm K}_0(\epsilon r) {\rm K}_0(\epsilon r^{\prime}) \left[ \sigma_{q\bar{q}}(z\rr,x) + \sigma_{q\bar{q}}(z\rp,x) - 
 \sigma_{q\bar{q}}(z|\rr - \rp|,x)\right] \nonumber \\
      & + & \frac{1}{2(2\pi)^2}\, \frac{({\cal C}^G_f)^2(g^{G}_{v,f})^2}{2\pi^2} \, [1+ (1-z)^2]\epsilon^2  \nonumber \\
      & \times & \int d^2\rr \dd^2\rp e^{i\pv \cdot (\rr - \rp)}  \frac{\rr\cdot\rp}{r r^{\prime}} {\rm K}_1(\epsilon r)
     {\rm K}_1(\epsilon r^{\prime})\,\left[ \sigma_{q\bar{q}}(z\rr,x) + \sigma_{q\bar{q}}(z\rp,x) - 
 \sigma_{q\bar{q}}(z|\rr - \rp|,x)\right] \,.
\end{eqnarray}     
Using the results presented in the Appendix B, we can derive that:
\begin{eqnarray}
z\frac{d\sigma^f_{T}}{dz d^2\pv}|_V & = &  \frac{({\cal C}^G_f)^2(g^{G}_{v,f})^2}{2\pi^2} \left\{ z^2[(m_b - m_a) + zm_a]^2 \, {\cal{D}}_1(z,p,\epsilon) +   \, [1+ (1-z)^2]\epsilon^2   \, {\cal{D}}_2(z,p,\epsilon) \right\}\,\,,
\label{Eq:VT_imp}
\end{eqnarray}
where we have defined the  auxiliary functions:
\begin{eqnarray}
{\cal{D}}_1(z,p,\epsilon) \equiv  \left[ \frac{1}{(p^2 + \epsilon^2)}I_1(z,p) - \frac{1}{4 \epsilon} I_2(z,p) \right] 
\end{eqnarray}
\begin{eqnarray}
{\cal{D}}_2(z,p,\epsilon) \equiv \left[\frac{1}{\epsilon} \frac{p}{(p^2 + \epsilon^2)}I_3(z,p) - \frac{1}{2\epsilon^2} I_1(z,p) +  \frac{1}{4\epsilon} I_2(z,p) \right] \,\,.
\end{eqnarray}

Similarly, substituting the expression for $\rho^A_V(z,\rr,\rr')$, one has that the axial contribution is given by: 
\begin{eqnarray}
z\frac{d\sigma^f_{T}}{dz d^2\pv}|_A & = & \frac{({\cal C}^G_f)^2(g^{G}_{a,f})^2}{2\pi^2} \left\{z^2[(m_b + m_a) - zm_a]^2 {\cal{D}}_1(z,p,\epsilon) + [1+ (1-z)^2]\epsilon^2   \, {\cal{D}}_2(z,p,\epsilon) \right\}
\label{Eq:AT_imp}
\end{eqnarray}

Following similar steps for the longitudinal polarization, results that
the vector contribution is given by
\begin{eqnarray}
z\frac{d\sigma^f_{L}}{dz d^2\pv}|_V & = &  \frac{({\cal C}^G_f)^2(g_{v,f}^G)^2}{4 \pi^2}\left\{\frac{(z^2 m_a(m_b - m_a) - z (m_b^2 - m_a^2) - 2 (1-z) M_G^2)^2}{ M_G^2}{\cal{D}}_1(z,p,\epsilon) \right. \nonumber \\
& \, & \left.  +  \frac{z^2(m_b-m_a)^2}{M_G^2} \epsilon^2 {\cal{D}}_2(z,p,\epsilon) \right\}
\label{Eq:VL_imp}
\end{eqnarray}
while the axial contribution can be written as
\begin{eqnarray}
z\frac{d\sigma^f_{L}}{dz d^2\pv}|_A & = &  \frac{({\cal C}^G_f)^2(g^{G}_{a,f})^2}{4 \pi^2} \left\{ \frac{(z^2 m_a(m_b + m_a) + z (m_b^2 - m_a^2) + 2 (1-z) M_G^2)^2}{M_G^2} {\cal{D}}_1(z,p,\epsilon)  +  \frac{\epsilon^2 z^2 (m_b+m_a)^2}{M_G^2} {\cal{D}}_2(z,p,\epsilon) \right\} \nonumber \\
\label{Eq:AL_imp}
\end{eqnarray}

\subsection{Representation in the transverse momentum space}
\label{subsec:mom}
Using the relation between the dipole - hadron cross-section and the unintegrated gluon distribution, given in Eq. (\ref{Eq:unint}), 
 one has that the differential cross-section will be given by
\begin{eqnarray}
\label{ptdistcc}
\frac{d\sigma^f_{T,L}}{dz d^2\pv} & = & \frac{1}{2(2\pi)^2}\, \overline{\sum}_\text{quark pol.}
\int d^2\rr \dd^2\rp e^{i\pv \cdot (\rr - \rp)}\, \Psi_{T,L}(z,\rr) 
\Psi^{ *}_{T,L}(z,\rp) \left[ \sigma_{q\bar{q}}(z\rr,x) + \sigma_{q\bar{q}}(z\rp,x) - 
 \sigma_{q\bar{q}}(z|\rr - \rp|,x)\right]\,, \nonumber \\
 & = & \frac{1}{2(2\pi)^2}\, \overline{\sum}_\text{quark pol.}
\int d^2\rr \dd^2\rp e^{i\pv \cdot (\rr - \rp)}\, \Psi_{T,L}(z,\rr) 
\Psi^{*}_{T,L}(z,\rp)  \int d^2\kv f(x,\kv)\,[1 - e^{i z \kv \cdot \rr} - e^{i z \kv \cdot \rp} + e^{i z \kv \cdot (\rr-\rp)}] \,\,. \nonumber \\
\end{eqnarray}

For the transverse polarization one has that
\begin{eqnarray}
\frac{d\sigma^f_{T}}{dz d^2\pv} & = & \frac{1}{2(2\pi)^2}\,\int d^2\rr \dd^2\rp e^{i\pv \cdot (\rr - \rp)}
\rho^T_V(z,\rr,\rr') \int d^2\kv f(x,\kv)\,[1 - e^{i z \kv \cdot \rr} - e^{i z \kv \cdot \rp} + e^{i z \kv \cdot (\rr-\rp)}] \nonumber \\
& + & \frac{1}{2(2\pi)^2}\,\int d^2\rr \dd^2\rp e^{i\pv \cdot (\rr - \rp)}
\rho^T_A(z,\rr,\rr') \int d^2\kv f(x,\kv)\,[1 - e^{i z \kv \cdot \rr} - e^{i z \kv \cdot \rp} + e^{i z \kv \cdot (\rr-\rp)}]
\end{eqnarray}
Substituting the expression for the squared LCWF, $\rho^T_V$, and using the results presented in the Appendix C, we obtain that the vector contribution can be expressed as follows:
\begin{eqnarray}
z\frac{d\sigma^f_{T}}{dz d^2\pv}|_V & = &  \frac{({\cal C}^G_f)^2(g^{G}_{v,f})^2}{2\pi^2}  \int d^2\kv f(x,\kv) \left\{ z^2[(m_b - m_a) + zm_a]^2 \, {\cal{E}}_1(\pv,\kv,\epsilon,z) + [1+ (1-z)^2] {\cal{E}}_2(\pv,\kv,\epsilon,z) \right\}\,\,,
\label{Eq:VT_trans}
\end{eqnarray}
where we have defined the  auxiliary functions:
\begin{eqnarray}
{\cal{E}}_1(\pv,\kv,\epsilon,z) &\equiv& \left[\frac{1}{2}\,\frac{1}{(p^2 + \epsilon^2)^2} -  \frac{1}{[(\pv-z \kv)^2 + \epsilon^2]} \frac{1}{(p^2 + \epsilon^2)} + 
\frac{1}{2} \frac{1}{[(\pv-z \kv)^2 + \epsilon^2]^2} \right] \nonumber \\
&=& \frac{1}{2} \left[ \frac{1}{p^2 + \epsilon^2} - \frac{1}{(\pv-z \kv)^2 + \epsilon^2}  \right]^2
\end{eqnarray}
\begin{eqnarray}
{\cal{E}}_2(\pv,\kv,\epsilon,z) &\equiv& \left[\frac{1}{2} \frac{p^2}{(p^2 + \epsilon^2)^2} -  \frac{\pv \cdot (\pv - z \kv)}{(p^2 + \epsilon^2)[(\pv-z \kv)^2 + \epsilon^2]}  + \frac{1}{2} \frac{(\pv - z \kv)^2}{[(\pv-z \kv)^2 + \epsilon^2]^2} \right] \nonumber \\
&=& \frac{1}{2} \left[ \frac{\pv}{p^2 + \epsilon^2} - \frac{\pv - z \kv}{(\pv-z \kv)^2 + \epsilon^2}  \right]^2
\end{eqnarray}
Similarly, substituting the expression for  $\rho^T_A$, one has that the axial contribution is given by: 
\begin{eqnarray}
z\frac{d\sigma^f_{T}}{dz d^2\pv}|_A & = &  \frac{({\cal C}^G_f)^2(g^{G}_{a,f})^2}{2\pi^2}  \,\int d^2\kv f(x,\kv)\left\{ z^2[(m_b + m_a) - zm_a]^2 {\cal{E}}_1(\pv,\kv,\epsilon,z) + [1+ (1-z)^2]{\cal{E}}_2(\pv,\kv,\epsilon,z) \right\} 
\label{Eq:AT_trans}
\end{eqnarray}

Following similar steps, one can derive that the vector and axial contributions for the longitudinal cross-section will be given by
\begin{eqnarray}
z\frac{d\sigma^f_{L}}{dz d^2\pv}|_V & = &  \frac{({\cal C}^G_f)^2(g_{v,f}^G)^2}{4\pi^2} \,\int d^2\kv f(x,\kv) \left\{\frac{(z^2 m_a(m_b - m_a) - z (m_b^2 - m_a^2) - 2 (1-z) M_G^2)^2}{ M_G^2} {\cal{E}}_1(\pv,\kv,\epsilon,z) \right. \nonumber \\
& \, & \left. +  \frac{z^2(m_b-m_a)^2}{M_G^2} {\cal{E}}_2(\pv,\kv,\epsilon,z) \right\} 
\label{Eq:VL_trans}
\end{eqnarray}
and
\begin{eqnarray}
z\frac{d\sigma^f_{L}}{dz d^2\pv}|_A & = &  \frac{({\cal C}^G_f)^2(g^{G}_{a,f})^2}{4\pi^2}   \,\int d^2\kv f(x,\kv) \left\{ \frac{(z^2 m_a(m_b + m_a) + z (m_b^2 - m_a^2) + 2 (1-z) M_G^2)^2}{M_G^2} 
{\cal{E}}_1(\pv,\kv,\epsilon) \right. \nonumber \\
& \,& \left. +  \frac{z^2 (m_b+m_a)^2}{M_G^2}  {\cal{E}}_2(\pv,\kv,\epsilon,z) \right\} 
\label{Eq:AL_trans}
\end{eqnarray}

The Eqs. (\ref{Eq:VT_imp}), (\ref{Eq:AT_imp}), (\ref{Eq:VL_imp}), (\ref{Eq:AL_imp}), (\ref{Eq:VT_trans}), (\ref{Eq:AT_trans}), (\ref{Eq:VL_trans}) and  (\ref{Eq:AL_trans}) are the main results of our paper. They allow us to estimate the vector and axial contributions for the spectrum associated with the isolated gauge boson production in the impact parameter and transverse momentum spaces. The expressions are valid for massive ($Z^0$ and $W^{\pm}$) and massless ($\gamma$) electroweak gauge bosons and take into account of the transverse and longitudinal polarizations as well as does not disregard the masses of the quarks, which is fundamental to derive realistic predictions for the $W^{\pm}$ production. 
In a forthcoming publication, we will perform  a detailed phenomenological study, where we will compare the resulting predictions with the current LHC data for the forward gauge boson production. However, as we will demonstrate in the next section, some particular cases of our expressions already have been used in the literature, with a successful description of the current data.

\section{Particular cases}
\label{sec:particular}

\subsection{Real photon production}
The simplest case that we can apply the expressions derived in the previous section is the real photon production in the $q p \rightarrow \gamma X$ process. For the production of a real photon, one has that $g_{a,f}^{\gamma}$ = 0,  $g_{v,f}^{\gamma}$ = 1,   $C_f^{\gamma} = \sqrt{\alpha_{em}} e_f$, $M_G^2 = 0$ and $m_a = m_b = m_f$. Moreover,  the longitudinal polarization does not contribute. Consequently, the differential cross-section  will be given in the impact parameter space by
\begin{eqnarray}
\left. z\frac{d\sigma^f_{T}}{dz d^2\pv} \right|_{q p \rightarrow \gamma X} & = &  \frac{\alpha_{em} e_f^2}{2\pi^2} \left\{ m_f^2 z^4 \, {\cal{D}}_1(z,p,\epsilon) +   \, [1+ (1-z)^2]\epsilon^2   \, {\cal{D}}_2(z,p,\epsilon) \right\}
\end{eqnarray}
with $\epsilon^2 =  z^2 m_f^2$. Such an expression was used as input in the calculations of the cross-section for the real photon production in $pp$ collisions performed e.g. in Refs. \cite{Kopeliovich:2007yva,Kopeliovich:2009yw,SampaiodosSantos:2020lte} using the color dipole formalism. On the other hand, in the transverse momentum space, the differential cross-sections will be given by:
\begin{eqnarray}
\left. z\frac{d\sigma^f_{T}}{dz d^2\pv} \right|_{q p \rightarrow \gamma X} & = &  \frac{\alpha_{em} e_f^2}{2\pi^2}  \int d^2\kv f(x,\kv) \left\{m_f^2 z^4 \, {\cal{E}}_1(\pv,\kv,\epsilon,z) + [1+ (1-z)^2] {\cal{E}}_2(\pv,\kv,\epsilon,z) \right\}\,.
\end{eqnarray}
Disregarding the quark mass, such an expression reduces to those used e.g. in  Refs. \cite{Gelis:2002ki,Jalilian-Marian:2012wwi,Ducloue:2017kkq,Goncalves:2020tvh,Lima:2023dqw} considering the CGC formalism to estimate the isolated photon production.

\subsection{Drell - Yan process}
Another possible application of our results is the emission of an off-mass shell gauge boson $G^*$ that decays into a dilepton system with invariant mass $M$,  characteristic of the Drell - Yan (DY) process. For $G^*=\gamma^*$ and $Z^*$, one has the production of $l^+ l^-$ system in the final state and such a process was analyzed in Refs. \cite{Kopeliovich:2000fb,Raufeisen:2002zp,Kopeliovich:2001hf,Betemps:2004xr,Betemps:2003je,Golec-Biernat:2010dup,Ducati:2013cga,Schafer:2016qmk,Ducloue:2017zfd,Gelis:2002fw,Baier:2004tj,Stasto:2012ru,Kang:2012vm,Basso:2016ulb}   using the color dipole and CGC formalisms.  The results derived in the previous section can be directly applied for this case, if we take into account that the decay process can be factorized of the gauge boson production as follows
\begin{eqnarray} \label{dilep}
\frac{d \sigma (qp\rightarrow [G^*\to l\bar l] X)}{dz d^2 \pv dM^2}=
{\cal F}_G(M)\,\frac{d \sigma (q p \rightarrow G^* X)}{dz d^2\pv} \,,
\end{eqnarray}
where the function ${\cal F}_G(M)$ describes the $G^* \rightarrow l^+ l^-$ and is given by 
\begin{eqnarray}
{\cal F}_\gamma(M) & = & \frac{\alpha_{em}}{3\pi M^2} \,   \hspace{2.9cm}  \mbox{for} \,\,\,  G^* = \gamma^*  \\
{\cal F}_Z(M) & = & \mathrm{Br}(Z^0\to l\bar l)\rho_Z(M) \,    \hspace{1cm}  \mbox{for} \,\,\,  G^* = Z^* \,\,,
\end{eqnarray}
where  $\rho_Z(M)$ is the invariant mass distribution of the $Z^0$ boson
in the narrow width approximation \cite{Kuksa:2008kw}
\begin{eqnarray}
\rho_Z(M)=\frac{1}{\pi}\, \frac{M\Gamma_Z(M)}{(M^2-m_Z^2)^2+[M\Gamma_Z(M)]^2}\,, \qquad \Gamma_Z(M)/M\ll 1 \,,
\end{eqnarray}
with generalized total $Z^0$ decay width being given by
\begin{eqnarray}
\Gamma_Z(M)=\frac{\alpha_{em}M}{6\sin^22\theta_W}\Big(\frac{160}{3}\sin^4\theta_W - 40 \sin^2\theta_W + 21\Big) \,,
\end{eqnarray}
where $\theta_W$ is the Weinberg gauge boson mixing angle in the SM.

For the particular case of a virtual photon ($G^*=\gamma^*$) with virtuality $M^2$, one has that the axial contributions vanishes and differential cross-section for the $q p \rightarrow [\gamma^* \rightarrow l\bar l] X$ process will be given in the impact parameter space by
\begin{eqnarray}
\left. z\frac{d\sigma^f_{total}}{dz d^2\pv dM^2}\right|_{DY}  & = & \left. z\frac{d\sigma^f_{T}}{dz d^2\pv dM^2}\right|_V + \left.z\frac{d\sigma^f_{L}}{dz d^2\pv dM^2}\right|_V \nonumber \\ & = &  \frac{\alpha_{em} e_f^2}{2\pi^2} {\cal F}_\gamma(M) \left\{[m_f^2 z^4 + 2M^2 \left(1-z\right)^2] \, {\cal{D}}_1(z,p,\epsilon) +   \, [1+ (1-z)^2]\epsilon^2   \, {\cal{D}}_2(z,p,\epsilon)   \right\}
\end{eqnarray}
with $\epsilon^2 = (1-z)M^2 + z^2 m_f^2$. Such an expression was used e.g. in Ref. \cite{Kopeliovich:2000fb,Raufeisen:2002zp,Kopeliovich:2001hf,Betemps:2004xr,Betemps:2003je,Golec-Biernat:2010dup,Ducati:2013cga,Schafer:2016qmk,Ducloue:2017zfd} to estimate the DY production in hadronic collisions at the LHC, assuming different models for the dipole - target cross-section. On the other hand, in the transverse momentum space, the differential cross-sections will be given by:
\begin{eqnarray}
\left. z\frac{d\sigma^f_{total}}{dz d^2\pv dM^2}\right|_{DY}  & = & \,\,
  \frac{\alpha_{em} e_f^2}{2\pi^2} {\cal F}_\gamma(M)  \int d^2\kv f(x,\kv) \left\{ [m_f^2 z^4 + 2M^2(1-z)^2]\, {\cal{E}}_1(\pv,\kv,\epsilon,z) + [1+ (1-z)^2] {\cal{E}}_2(\pv,\kv,\epsilon,z) \right\}\,\,, \nonumber \\
  \end{eqnarray}
    which was derived in the massless limit in Ref. \cite{Gelis:2002fw} using the CGC formalism and applied for $pp/pA$ collisions at the LHC, e.g., in Refs. \cite{Baier:2004tj,Stasto:2012ru,Kang:2012vm,Basso:2016ulb}. 

Finally, it is important to emphasize that we have verified that our results also reproduce the expressions presented in Ref. \cite{Basso:2015pba} for the $G^* = Z^*$ case.

\subsection{The color transparency regime}
\label{sec:r2approx}

The behavior of the spectrum for the electroweak gauge boson production is 
strongly dependent on the dipole - proton cross - section, which is 
determined by the QCD dynamics at high energies. In recent years, several 
groups have proposed distinct models for this quantity, some derived from 
the solutions of the Balitsky - Kovchegov equation \cite{Albacete:2009fh,Albacete:2010sy} or inspired by its 
solutions in the saturation and linear  regimes \cite{iim}, and others based on 
a particular phenomenological ansatz \cite{kt,kmw}. In general, the dependence on $\rr^2$ is non-trivial and energy dependent, which makes the calculation of the auxiliary functions ${\cal{D}}_i$'s  present in the spectrum a hard task due to the presence of oscillatory functions in its definitions. In a forthcoming publication, we will present the results of our numerical calculations of the $pp$ cross-sections performed considering the most recent models for $\sigma_{q\bar{q}}$. Here we will analyze in more detail the color transparency regime, predicted to be valid when $\rr \rightarrow 0$, which implies that the dipole - proton cross-section can be approximated by $\sigma_{q\bar{q}} \propto \rr^2$. Such behavior is also denoted in the literature by $\rr^2$ -- approximation, and  is expected to be valid when the impact of the saturation corrections is negligible. As we will demonstrate below, in this regime, it is possible to derive analytical expressions for the spectrum, which can be used to study the $z$ and $\pv$ dependencies in the linear regime and can be considered as a baseline for future comparisons with the results derived using more sophisticated models for the dipole - proton cross-section.

In order to perform our calculations, we will consider the linear prediction of the GBW model \cite{GBW}, in which $\sigma_{q\bar{q}} (\rr,x) = {\cal{C}}(x) \rr^2$ with ${\cal{C}}(x) = \sigma_0 Q_s^2(x)/4$, where $Q_s = (x_0/x)^{\lambda/2}$ is the saturation scale and the parameters $\sigma_0$, $x_0$ and $\lambda$ are determined by a fit to the HERA data. In this approximation the $r$ - integrals in the functions $I_1$, $I_2$ and $I_3$ can be performed analytically and are given by 
\begin{eqnarray}
 I_1(z,p) & = &  \int dr r {\rm J}_0(pr) {\rm K}_0 (\epsilon r) \sigma_{q\bar{q}}(z\rr)  = {\cal{C}}(x) \frac{4z^2(\epsilon^2 - p^2)}{(p^2 + \epsilon^2)^3}
\end{eqnarray}
\begin{eqnarray}
 I_2(z,p) & = &  \int dr r^2  {\rm J}_0(pr) {\rm K}_1 (\epsilon r) \sigma_{q\bar{q}}(z\rr)  = {\cal{C}}(x) \frac{16 z^2 \epsilon (\epsilon^2  - 2p^2)}{(p^2 + \epsilon^2)^4} 
\end{eqnarray}
\begin{eqnarray}
I_3(z,p) =  \int dr r  {\rm J}_1(pr) {\rm K}_1 (\epsilon r) \sigma_{q\bar{q}}(z\rr)  = {\cal{C}}(x) \frac{8 z^2 \epsilon p}{(p^2 + \epsilon^2)^3} 
\end{eqnarray}
As a consequence, the auxiliary functions ${\cal{D}}_1(z,p,\epsilon) $ and ${\cal{D}}_2(z,p,\epsilon)$ can be expressed as follows 
\begin{eqnarray}
  {\cal{D}}_1(z,p,\epsilon) &=&  \frac{1}{(p^2 + \epsilon^2)}{\cal{C}}(x) \frac{4z^2(\epsilon^2 - p^2)}{(p^2 + \epsilon^2)^3} - \frac{1}{4 \epsilon} {\cal{C}}(x) \frac{16 z^2 \epsilon (\epsilon^2  - 2p^2)}{(p^2 + \epsilon^2)^4}   \\
  {\cal{D}}_2(z,p,\epsilon) &=& \frac{1}{\epsilon} \frac{p}{(p^2 + \epsilon^2)}{\cal{C}}(x) \frac{8 z^2 \epsilon p}{(p^2 + \epsilon^2)^3}  
  - \frac{1}{2\epsilon^2} {\cal{C}}(x) \frac{4z^2(\epsilon^2 - p^2)}{(p^2 + \epsilon^2)^3} 
  +  \frac{1}{4\epsilon} {\cal{C}}(x) \frac{16 z^2 \epsilon (\epsilon^2  - 2p^2)}{(p^2 + \epsilon^2)^4}  \,\,,
\end{eqnarray}
which can be simplified as 
\begin{eqnarray}
  \mathcal D_1(z,p,\epsilon) &=& {\cal{C}}(x) \frac{4z^2 p^2}{(p^2+\epsilon^2)^4} \\ 
  \mathcal D_2(z,p,\epsilon) &=& {\cal{C}}(x) \frac{2z^2\epsilon^2}{(p^2+\epsilon^2)^4}
  \left[ 1 + \frac{p^4}{\epsilon^4} \right] \,\,.
\end{eqnarray}
Such expressions can be used in Eqs. (\ref{Eq:VT_imp}), (\ref{Eq:AT_imp}), (\ref{Eq:VL_imp}) and (\ref{Eq:AL_imp}) to obtain the distinct contributions for the differential cross-section in the color transparency regime.



\section{Summary}
\label{conc}
The description of the particle production at forward rapidities in proton - proton and proton - nucleus collisions at high energies is still one of the main challenges of the strong interactions theory. In this kinematical region, new effects associated with the non-linear effects in the QCD dynamics are expected to contribute and the usual treatment of the cross-section in terms of the collinear factorization formalism is predicted to breakdown. Over the last decades, several authors have discussed possible generalized factorization formalisms as well as improved the description of saturation effects in the hadronic wave functions. In general, the final formulas for the cross-sections associated with the production of two partons/hadrons  involve new quantities, which are non-trivial to estimate. In this paper, we addressed a simpler process, where one of the particles in the final state is an electroweak gauge boson $G$. As we have demonstrated, for this case,  the differential cross-section for the associated production of the gauge boson with a quark can be fully expressed in terms of the squared light cone wave function (LCWF) for the $q_f \rightarrow G q_k$ transition, and the  usual dipole - proton cross - section, which can be constrained by the HERA data. In the current study, we have derived, for the first time, the generic expressions for the LCWF's. Moreover, we have estimated the vector and axial contributions for the description of the longitudinal and transverse spectra associated with the isolated gauge boson production in the impact parameter and transverse momentum spaces. In addition, we demonstrated that our results reduce to expressions previously used in the literature for the description of the real photon production and Drell - Yan process at forward rapidities in some particular limits. Finally, analytical expressions for the spectrum have been derived in the color transparency limit, which is a reasonable approximation of the linear regime of QCD dynamics. The results derived in this paper are the main  ingredients for the calculation of the $pp$ cross - sections, which can be compared with the current and forthcoming LHC data. Such studies are being performed and will be presented in separated publications.

\section*{Acknowledgments}
Y.B.B.  and V.P.G. were  partially supported by CNPq, CAPES (Finance code 001), FAPERGS and  INCT-FNA (Process No. 464898/2014-5). 

\appendix

\section{Bessel Functions}

\begin{eqnarray}
{\rm J}_0(p r)=\frac{1}{2\pi}\int d\phi e^{i \pv \cdot \rr}=\frac{1}{2\pi}\int d\phi e^{i p r\cos\phi}.
\end{eqnarray}

\begin{eqnarray}
{\rm J}_1(p r)=\frac{1}{2\pi i} \int d\phi e^{i p r\cos\phi} \cos\phi.
\end{eqnarray}

\begin{eqnarray}
{\rm K}_0(\varepsilon r) = \frac{1}{2 \pi}\int d^2\lv e^{i \lv \cdot \rr} \frac{1}{(l^2+\epsilon^2)}
\end{eqnarray}

\begin{eqnarray}
{\rm K}_1(\varepsilon r) = - \frac{1}{\epsilon} \frac{d}{dr} {\rm K}_0 (\epsilon r)  =   \frac{1}{2 \pi i \epsilon}\int d^2\lv e^{i \lv \cdot \rr} \frac{\lv}{(l^2+\epsilon^2)} \cdot   \frac{\rr}{r} 
\end{eqnarray}
Consequently:
\begin{eqnarray}
\frac{\rr}{r} {\rm K}_1(\varepsilon r) = \frac{1}{2 \pi i \epsilon}\int d^2\lv e^{i \lv \cdot \rr} \frac{ \lv}{(l^2+\epsilon^2)}
\end{eqnarray}
Moreover, we can also write:
\begin{eqnarray}
{\rm K}_1(\varepsilon r) = - \frac{1}{r} \frac{d}{d\epsilon} {\rm K}_0 (\epsilon r) =   \frac{\epsilon}{ \pi r}\int d^2\lv e^{i \lv \cdot \rr} \frac{1}{(l^2+\epsilon^2)^2} 
\end{eqnarray}

\section{Useful integrals for the calculations in the impact parameter space}

\begin{eqnarray}
\int d^2\rr d^2\rp \exp[i \pv \cdot (\rr - \rp)] {\rm K}_0(\epsilon r) {\rm K}_0(\epsilon r^{\prime}) \sigma_{q\bar{q}}(z\rr) & = &  \frac{(2 \pi)^2}{(p^2 + \epsilon^2)}  \int dr r  {\rm J}_0(pr) {\rm K}_0 (\epsilon r) \sigma(z\rr) \nonumber \\ & = & \frac{(2 \pi)^2}{(p^2 + \epsilon^2)}  I_1(z,p)
\end{eqnarray}


\begin{eqnarray}
\int d^2\rr d^2\rp \exp[i \pv \cdot (\rr - \rp)] {\rm K}_0(\epsilon r) {\rm K}_0(\epsilon r^{\prime}) \sigma_{q\bar{q}}(z|\rr - \rp|) & = & \frac{2\pi^2}{\epsilon}  \int dr r^2  {\rm J}_0(pr) {\rm K}_1 (\epsilon r) \sigma(z\rr) \nonumber \\ & = & \frac{2\pi^2}{\epsilon}   I_2(z,p)
\end{eqnarray}

\begin{eqnarray}
\int d^2\rr d^2\rp \exp[i \pv \cdot (\rr - \rp)] \frac{\rr \cdot \rp}{r r^{\prime}} {\rm K}_1(\epsilon r) {\rm K}_1(\epsilon r^{\prime}) \sigma_{q\bar{q}}(z\rr) & = &  \frac{(2 \pi)^2}{\epsilon} \frac{p}{(p^2 + \epsilon^2)}  \int dr r  {\rm J}_1(pr) {\rm K}_1 (\epsilon r) \sigma_{q\bar{q}}(z\rr) \nonumber \\ & = & \frac{(2 \pi)^2}{\epsilon} \frac{p}{(p^2 + \epsilon^2)} I_3(z,p)
\end{eqnarray}


\begin{eqnarray}
\int d^2\rr d^2\rp \exp[i \pv \cdot (\rr - \rp)] \frac{\rr \cdot \rp}{r r^{\prime}} {\rm K}_1(\epsilon r) {\rm K}_1(\epsilon r^{\prime}) \sigma_{q\bar{q}}(z|\rr - \rp|) & = &  \frac{(2 \pi)^2}{\epsilon^2} I_1(z,p) - \frac{2 \pi^2}{\epsilon} I_2(z,p)
\end{eqnarray}

\section{Useful integrals for the calculations in the transverse momentum space}

\begin{eqnarray}
\int d^2\rr d^2\rp \exp[i \pv \cdot (\rr - \rp)] {\rm K}_0(\epsilon r) {\rm K}_0(\epsilon r^{\prime}) = (2 \pi)^2 \frac{1}{(p^2 + \epsilon^2)^2}  
\end{eqnarray}

\begin{eqnarray}
\int d^2\rr d^2\rp \exp[i \pv \cdot (\rr - \rp)] {\rm K}_0(\epsilon r) {\rm K}_0(\epsilon r^{\prime}) e^{i z \kv \cdot \rr} = (2 \pi)^2 \frac{1}{[(\pv - z \kv)^2 + \epsilon^2]} \frac{1}{(p^2 + \epsilon^2)}  
\end{eqnarray}


\begin{eqnarray}
\int d^2\rr d^2\rp \exp[i \pv \cdot (\rr - \rp)] {\rm K}_0(\epsilon r) {\rm K}_0(\epsilon r^{\prime}) e^{i z \kv \cdot (\rr - \rp)} = (2 \pi)^2 \frac{1}{[(\pv - z\kv)^2 + \epsilon^2]^2}   
\end{eqnarray}

\begin{eqnarray}
\int d^2\rr d^2\rp \exp[i \pv \cdot (\rr - \rp)] \frac{\rr \cdot \rp}{r r^{\prime}} {\rm K}_1(\epsilon r) {\rm K}_1(\epsilon r^{\prime}) =  \frac{(2 \pi)^2}{\epsilon^2} \frac{p^2}{(p^2 + \epsilon^2)^2}    
\end{eqnarray}

\begin{eqnarray}
\int d^2\rr d^2\rp \exp[i \pv \cdot (\rr - \rp)] \frac{\rr \cdot \rp}{r r^{\prime}} {\rm K}_1(\epsilon r) {\rm K}_1(\epsilon r^{\prime}) e^{i z \kv \cdot \rr} = \frac{(2 \pi)^2}{\epsilon^2} \frac{\pv \cdot (\pv - z \kv)}{(p^2 + \epsilon^2) [(\pv - z \kv)^2 + \epsilon^2]}   
\end{eqnarray}


\begin{eqnarray}
\int d^2\rr d^2\rp \exp[i \pv \cdot (\rr - \rp)] \frac{\rr \cdot \rp}{r r^{\prime}} {\rm K}_1(\epsilon r) {\rm K}_1(\epsilon r^{\prime}) e^{i z \kv \cdot (\rr - \rp)} = \frac{(2 \pi)^2}{\epsilon^2} \frac{(\pv - z \kv)^2}{[(\pv - z \kv)^2 + \epsilon^2]^2}   
\end{eqnarray}

\section{Spinor matrix elements}
\label{ap:spinor}

Here we collect the spinor matrix elements relevant for the calculation of LCWFs. We use the spinors of Lepage and Brodsky \cite{Lepage:1980fj}, adjusted for the fact, that we define LF components as $p^\pm = (p^0 \pm p^3)/\sqrt{2}$. Notice that in general the spinors for initial and final state quark can belong to different masses.

\begin{eqnarray}
    \overline u(p_b,\lambda',m_b)\openone u_\lambda(p_a,\lambda,m_a) &=& 
    \sqrt{p_b^+p_a^+} \, \chi^\dagger_{\lambda'} 
    \left\{
        \left( 
            \frac{m_a}{p_a^+} + \frac{m_b}{p_b^+}
        \right)\openone
        - \frac{\pmb\sigma\cdot\pmb p_b}{p_b^+} (\pmb \sigma \cdot \pmb n)
    \right\}\chi_\lambda \nonumber \\
   \overline u(p_b,\lambda',m_b) \gamma_5 u(p_a,\lambda,m_a)&=&
    \sqrt{p_a^+ p_b^+} \chi^\dagger_{\lambda'}
    \left\{ 
        \left(
            \frac{m_b}{p_b^+} - \frac{m_a}{p_a^+}
        \right) (\pmb \sigma \cdot \pmb n)
        - \frac{\pmb\sigma\cdot\pmb p_b}{p_b^+}
    \right\}\chi_\lambda 
    \end{eqnarray}
For completeness, we give all components of the vector and axial vector bilinears, although the LF minus component does not appear in the gauge adopted in this paper. 
For $j=[j^+,j^-,\bj]$, we obtain for the vector and axial vector bilinear
\begin{eqnarray}
    j^V_\mu = \bar u(p_b,\lambda',m_b) \gamma_\mu u(p_a,\lambda,m_a) \, , \, j^A_\mu = \bar u(p_b,\lambda',m_b) \gamma_\mu \gamma_5 u(p_a,\lambda,m_a) \, , 
\end{eqnarray}
are given, respectively, by
\begin{eqnarray}
j^{V+}_{\lambda' \lambda} &=& 2 \sqrt{p_b^+ p_a^+} \,\, \chi^\dagger_{\lambda'} \openone \chi_\lambda \,  \nonumber \\
j^{V-}_{\lambda' \lambda} &=& \sqrt{p_b^+ p_a^+} \, \,  \chi^\dagger_{\lambda'} \Big\{ \frac{m_a m_b}{p_a^+ p_b^+} \openone  - \frac{m_a}{p_a^+ p_b^+} (\bsigma \cdot \pv_b) (\bsigma
\cdot \nv) \Big\} \chi_\lambda \nonumber \\
\bj^V_{\lambda' \lambda} \cdot \ba  &=& \sqrt{p_a^+ p_b^+} \, \, \chi^\dagger_{\lambda'} \Big\{ \Big( \frac{\pv_b \cdot \ba}{p_b^+} + i  \frac{[\pv_b, \ba]}{p_b^+} (\bsigma \cdot \nv) \Big) \openone + \frac{p_a^+ m_b - p_b^+ m_a}{p_a^+ p_b^+} \, (\bsigma\cdot \ba)(\bsigma \cdot \nv) \Big\} \chi_\lambda  \, ,
\end{eqnarray}
and
\begin{eqnarray}
j^{A+}_{\lambda ' \lambda}
 &=& 2 \sqrt{p_a^+ p_b^+} \,  \chi^\dagger_{\lambda'} (\bsigma \cdot \nv) \chi_\lambda
\nonumber \\
j^{A-}_{\lambda' \lambda} &=& \sqrt{p^+_a p^+_b}
\, \, \chi^\dagger_{\lambda'} \Big\{ -\frac{m_a m_b}{p_a^+ p_b^+} (\bsigma \cdot \nv) + \frac{m_a}{p_a^+ p_b^+} \, \bsigma \cdot \pv_b \Big\} \chi_\lambda
\nonumber \\
\bj^A_{\lambda' \lambda} \cdot \ba  &=& \sqrt{p^+_a p^+_b}
\, \, \chi^\dagger_{\lambda'} \Big\{ \Big( \frac{\pv_b \cdot \ba}{p_b^+} (\bsigma \cdot \nv) + i \frac{[\pv_b, \ba]}{p_b^+} \Big)  \openone + \frac{p_a^+ m_b + p_b^+ m_a}{p_a^+ p_b^+} \,  \bsigma\cdot \ba \Big\} \chi_\lambda \, .
\end{eqnarray}
Above, vectors $\pv_b$ is the transverse momentum of quark $b$, $\ba$ is an arbitrary transverse vector, and $\nv = (0,0,1)$. Pauli spinors $\chi_\lambda$ are eigenstates of $\bsigma \cdot \nv$:
\begin{eqnarray}
    (\bsigma \cdot \nv) \chi_\lambda = \lambda \, \chi_\lambda, \quad \lambda = \pm 1 ,
\end{eqnarray}
so that $\lambda/2$ is the quark helicity.

\bibliographystyle{unsrt}

\end{document}